\documentclass[onecolumn,aps,prb,final,amsfonts,amssymb,
amsmath,floatfix,showpacs]{revtex4}

\usepackage{graphicx}
\usepackage{bm}
\DeclareMathOperator{\im}{Im}
\DeclareMathOperator{\re}{Re}

\begin{document}

\title{Time resolved excitonic coherent polarization dynamics 
in GaAs multiple quantum wells}
\author{Bipul Pal}
\email{bipulpal@tifr.res.in}
\author{A.~S.~Vengurlekar}
\affiliation{Tata Institute of Fundamental Research, Mumbai 400005, India}
\date{\today}

\begin{abstract}
We investigate pump induced modulation in the temporal evolution
of the excitonic coherent polarization generated by a probe pulse
in GaAs multiple quantum wells. For this, we perform fs time
resolved pump-probe reflectivity measurements using frequency
upconversion at both positive and negative pump-probe delays.
A contribution arising from interference of excitonic linear
and nonlinear polarization, which is usually not considered in
describing pump-probe experiments, is required to understand
the results.
\end{abstract}


\maketitle

Linear and nonlinear optical response of excitonic resonances in
GaAs like semiconductor multiple quantum wells (MQW) to excitation 
by ultrashort laser pulses and the behavior of the primary and 
secondary radiation emitted by the excited macroscopic polarization 
has been of much interest in recent years. Measurement of time 
resolved emission in the reflection~\cite{weber,haacke,hayes} and 
transmission~\cite{kim,lyngnes} directions as a result of resonant 
excitation of excitons in quantum wells (QW) by an ultrashort laser 
pulse showed a  time-profile significantly different from that of 
the incident pulse as a result of coherent emission from the 
excitonic polarization surviving up to several ps. A modification 
in the temporal evolution of the coherent emission in the probe 
reflection and transmission direction occurs in pump-probe 
differential (PPD) measurements. This is caused by linear and 
nonlinear interactions among the populations and polarizations, 
excited in the MQW by the probe and the pump pulses, and the 
propagating electric fields. An interesting manifestation of 
the nonlinear optical effects in the pump-probe differential 
transmission~\cite{fluegel,joffre,sokoloff,neukirch} (PPDT) and 
reflection~\cite{guenther,bipul} (PPDR)  is the transient 
oscillations observed in spectrally resolved (SR) measurements for
negative ($-$ve) pump-probe delay ($\tau$). (Here, $-$ve delay 
corresponds to the situation when the probe pulse precedes the 
pump pulse.) As the PPD measurements entail detection of intensity  
either of the spectrally or time resolved signals, and as these 
intensities are not simply Fourier related, important phase related 
effects are not the same in the two measurements. Therefore, 
additional insights into the modification of the emission in the 
probe reflection by a pump pulse may be obtained from time domain 
measurements. However, to our knowledge, such measurements have 
not been reported so far.

The purpose of this paper is two-fold. First, we present our
experimental investigations of the time resolved PPDR from 
excitons in GaAs MQWs performed using the frequency upconversion 
technique. We find that the modulation in the dynamics of probe 
induced excitonic coherence by a pump pulse at low intensities 
occurs with a rise time of a few ps. However, the rise becomes 
faster as the pump intensity increases. No signal having the pump 
pulse polarization is emitted along the probe reflection direction 
for pump-probe cross-linear polarization. A PPDR signal is observed 
for $-$ve delay in both time resolved (TR) and time integrated 
(TI) measurements. The other main objective of this paper is to 
point out that the theoretical expressions usually employed to 
explain the PPD measurements~\cite{srink,shah} are inadequate in 
explaining the TR- and TI-PPDR measurements. Traditionally, the PPD 
signal is expressed as an interference of the probe electric field 
with the excitonic nonlinear polarization. This was able to 
explain the SR-PPDT results~\cite{lindberg,sokoloff,koch} in the 
regime of the third order nonlinear optical susceptibility. 
One prediction of this 
description is that no signal in TR-PPD measurements can occur 
for $\tau<0$. This theory allows nonzero signal only for $\tau>0$ 
and that too at the time of arrival of the probe pulse in a 
narrow time domain limited by the probe pulse width. Furthermore, 
the model also predicts that the TI-PPD (or equivalently, 
spectrally integrated (SI) PPD) signal at $-$ve delay should 
vanish. However, all these predictions are contradictory to our 
experimental results. To understand the origin and nature of 
the signals observed in TR- and TI-PPDR measurements, we 
reexamine the theory for calculating the PPD signals. We show 
that an additional nonlinear optical contribution arising from 
interference of excitonic linear and nonlinear polarization, 
which is usually ignored in the literature, has to be invoked 
to explain our results.

\begin{figure}[t]
\includegraphics[clip,width=6.5cm]{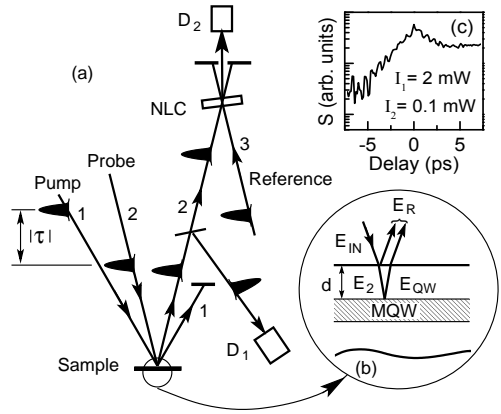}
\caption{ \label{expt}
(a) Schematics of the experimental arrangements. NLC~=~nonlinear 
crystal, D~=~photomultiplier tube detector. (b) The propagation 
directions for the incident ($E_{I\!N}$) and reflected ($E_{R}$) 
probe pulse are shown, together with a schematic picture of the 
sample having the MQW and a top layer (of thickness $d$). $E_{2}$ 
is the incident probe electric field on the QWs and $E_{QW}$ is the 
field emitted from the QWs. (c) The magnitude of the time (and 
spectrally) integrated PPDR signal as a function of pump-probe 
delay is plotted on a semilogarithmic scale. $I_{1}\ (I_{2})$ 
is the average intensity of the pump (probe) pulse.
}
\end{figure}

Our experiments are performed at 8~K on 17.5~nm thick GaAs
multiple-QWs, 20 in number, separated by 15~nm 
Al$_{0.33}$Ga$_{0.67}$As barriers, with a 330~nm 
AlGaAs layer on top. Continuous wave 
photoluminescence spectra measured at 8~K reveal heavy hole (hh) 
exciton emission at 1.5305~eV, with a spectral full width at half 
maximum (FWHM) 0.8~meV. Ti-Sapphire laser pulses of width of 
180~fs (spectral FWHM of 10~meV) and a repetition rate of 82~MHz 
are used to resonantly excite the hh excitons. However, as the hh 
and light hole (lh) exciton energies are separated by 5.7~meV, 
coherent excitation of lh excitons also occurs. Dependence of 
TI- and TR-PPDR signal on the pump-probe delay is measured using 
the experimental arrangement shown schematically in 
Fig.~\ref{expt}(a). The TR-PPDR signal as well as the time
profile of the reflected probe pulse in absence of a pump 
pulse are measured using type-I frequency upconversion in a 
LiIO$_{3}$ crystal on which the reflected probe pulse and a 
reference pulse are cofocussed. The pump, probe and reference 
pulses are all derived from the same laser beam. 
Figure~\ref{expt}(b) schematically shows the pulse propagation 
directions for the incident and reflected probe beam in the MQWs.

Delay dependence of the magnitude of the time (and spectrally) 
integrated PPDR signal (which is actually $-$ve) is 
plotted in  Fig.~\ref{expt}(c). Note that the TI-PPDR signal 
does not vanish at $-$ve delay up to about $-6$ ps. The signal 
decay for $\tau<0$ is nearly exponential with a time constant 
of about 2~ps. For large positive ($+$ve) delay, the signal 
decay is slow and is presumably governed by the exciton life time 
(full data is not shown). At small $+$ve delay, an additional 
fast decaying component is seen. The oscillations seen in the 
data at $-$ve and small $+$ve delay are the signature of quantum 
beats due to coexcitation of hh and lh excitons.

\begin{figure}[h]
\includegraphics[clip,width=7.5cm]{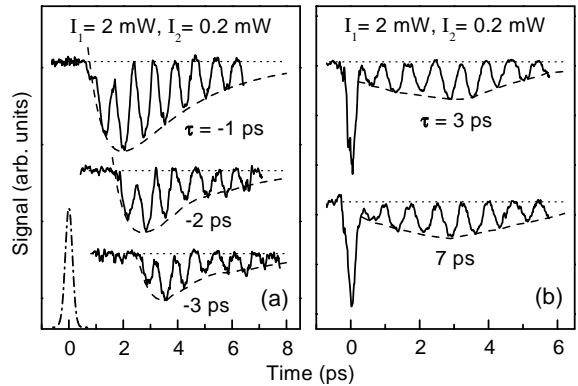}
\caption{ \label{delay}
Time evolution of the TR-PPDR signal (solid lines) at a few $-$ve
(a) and $+$ve (b) delays are shown on linear scales. Here pump 
(probe) intensity $I_{1}\ (I_{2})=2 \ (0.2)$~mW. The signal is 
$-$ve and the data for different $\tau$ are shifted along the 
vertical axis, with the zero of the signal shown as dotted lines. 
The dashed lines are shown as the envelope of the signal to guide 
the eye. The cross-correlation trace for the probe pulse is 
plotted in (a) (dashed-dotted line).
}
\end{figure}

\begin{figure}[h]
\includegraphics[clip,width=7.5cm]{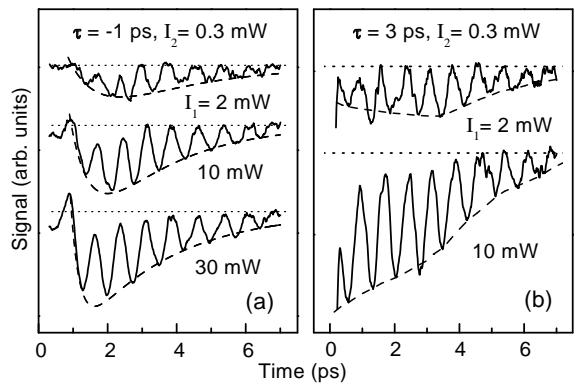}
\caption{ \label{int}
Time evolution of the TR-PPDR signal (solid lines) at a few pump 
intensities ($I_{1}$)  at $\tau=-1$~ps (a) and 3~ps (b), with 
$I_{2}=0.3$~mW. The signal is $-$ve and the data for different 
$\tau$ are shifted along the vertical axis (linear scale), with 
the zero of the signal shown as dotted lines. The dashed lines 
are shown as the envelope of the signal to guide the eye.
}
\end{figure}

The TR-PPDR measurements are shown in Figs.~\ref{delay} and 
\ref{int} for both $-$ve and $+$ve $\tau$. Here, we take the 
probe pulse to arrive at $t=0$ and the pump pulse at $t=-\tau$, 
so that the pump pulse arrives after (before) the probe pulse 
when $\tau<0$ ($\tau>0$). The average pump (probe) intensity 
$I_{1}$ ($I_{2}$) is 2~(0.2)~mW in Fig.~\ref{delay}. For the 
data in Fig.~\ref{int}, $I_{1}$ is varied, keeping $I_{2}$ 
fixed at 0.3~mW. The data shows quantum beat oscillations 
(Figs.~\ref{delay} and \ref{int}) with a period of about 
0.75~ps, corresponding closely with the hh-lh exciton energy 
separation of 5.7~meV. Also, shown in Fig.~\ref{delay}(a) 
(dash-dotted line) is the cross-correlation trace of the 
laser pulse, giving the time resolution of the upconversion 
measurement of about 250~fs. Note that the 
TR-PPDR signal in Figs.~\ref{delay} and \ref{int} is $-$ve at 
all $t$. This means that the coherent polarization induced by 
the probe pulse in the MQWs is reduced due to the influence of 
the pump pulse for both $-$ve and $+$ve delay. We show later 
that such a reduction may be explained by invoking effects 
like phase space filling and excitation induced dephasing 
within the framework of optical Bloch equations. (However, 
local field or exciton energy shift have no role in this.)

Consider first the case of $\tau<0$.  The emission along the 
probe reflection direction following excitation by the probe 
pulse may survive so long as the exciton polarization radiates 
coherently. A pump pulse incident during this process modifies 
the subsequent emission. Figure~\ref{delay}(a) shows the time
evolution of the PPDR signal measured at three $-$ve delays. 
As expected from causality, the onset of the TR-PPDR signal 
correlates with the arrival of the pump pulse, delayed with 
respect to the probe pulse. No signal is observed when $\tau$ 
exceeds about 8~ps as the probe pulse induced polarization 
dephases within this time. At low intensities, the signal shows 
a rise time larger than the laser pulse width. After reaching 
the peak, it decays approximately exponentially with a time 
constant of about 2.5~ps. Figure~\ref{int}(a) plots the TR-PPDR 
signal measured at three values of average pump intensity
$I_{1}$ at $\tau=-1$~ps. While the decay rate of the signal 
appears to be rather insensitive to $I_{1}$, a reduction in 
rise time with increased $I_{1}$ is clearly seen.

The observation of a nonzero TR-PPDR signal even when the pump pulse 
is incident on the sample well after the probe pulse has passed 
may be compared with a similar situation noted previously in the
case of four wave mixing (FWM) measurements~\cite{leo,wegener}. 
However, within the regime of the third order nonlinear optical 
susceptibility, the nonlinear optical polarization relevant to 
the PPD is not the same as that in FWM. 
For example, transient oscillations are seen in SR-PPD 
measurements at $-$ve delays, but not in SR-FWM. The exact origin 
of the signals at $-$ve $\tau$ is different in the two cases. In 
the FWM case, the polarization induced by the pulse along 
$\bm{\mathrm{k}}_{2}$ gets diffracted along 
$2\bm{\mathrm{k}}_{2}-\bm{\mathrm{k}}_{1}$ after the other 
excitation pulse arrives along $\bm{\mathrm{k}}_{1}$. This is  
a result of many body Coulomb interactions~\cite{wegener,wang,mayer} 
(such as local field, biexcitons, excitation induced dephasing, 
for example). On the other hand, the TR-PPDR signal at $\tau<0$ 
is in principle possible within the third order nonlinear optical 
regime even without introducing the many body Coulomb interactions, 
as shown presently.

It is sometime stated that the SR-PPD signal observed for 
$\tau<0$ is a result of pump pulse being scattered in the probe 
reflection/transmission direction from a grating created by the 
probe induced exciton polarization and the pump 
pulse~\cite{sokoloff,koch,shah}. If a transfer of energy from 
the pump beam to the probe reflection direction occurs, this may 
give rise to a $+$ve TR-PPDR signal, contrary to the data of 
Figs.~\ref{delay}(a) and \ref{int}(a). We further investigate 
this point by performing the TR-PPDR measurements for the pump 
polarization parallel and perpendicular to the probe (and the 
reference pulse) polarization. We find that the TR-PPDR signal 
which is measured using type-I upconversion is identical for 
the two cases of pump polarization. This confirms that the 
emission in the probe reflection direction always has the 
polarization of the probe pulse~\cite{sieh}.

Figure~\ref{delay}(b) displays the modification caused by 
a pump pulse in the time evolution of radiation from the 
excitonic polarization induced in the MQWs by a probe pulse 
incident after a $+$ve delay ($\tau>0$).  Unlike $\tau<0$, 
the TR-PPDR signal in this case has a large contribution at $t=0$, 
and then a tail for $t>0$. The signal at $t=0$ is essentially caused 
by the interference of the probe electric field with the excitonic 
nonlinear polarization induced by the pump and probe pulses. The 
signal in the tail part shows a rise and then a decay with $t$. 
The rise for $\tau>0$ [Fig.~\ref{delay}(b)] is slower compared to 
the $\tau<0$ case seen in Fig.~\ref{delay}(a). In fact, the overall 
shape of the $t$ dependence is different in the two cases 
[Figs.~\ref{delay}(a) and (b)]. Also, the TR-PPDR signal shows 
only a small dependence on the value of $\tau$ when $\tau$ is 
$+$ve. (We verify this by varying $\tau$ from 
1 to 30~ps.) The pump pulse excites its 
own coherent exciton polarization which turns into incoherent 
excitons after several ps. The rather small sensitivity of the 
TR-PPDR signal to change in $\tau$~($>0$) exhibited in 
Fig.~\ref{delay}(b) indicates that the modulation of emission in 
the probe reflection direction by coherent and incoherent excitons 
induced by the pump pulse is not strongly distinguishable. In 
Fig.~\ref{int}(b), the TR-PPDR signal at small $I_{1}$ shows an 
initial rise before decaying for larger $t$. As $I_{1}$ is 
increased, the rising part is not seen, the signal decaying 
continuously with $t$. If the decay of the coherent emission in 
the reflected probe pulse direction were exponential in $t$, and 
if the pump modulation only enhances the exponential decay rate, 
the TR-PPDR signal should be given simply as a difference of two 
exponentially decaying terms which may show an initial rise and 
then a decay. We find that the data of Fig.~\ref{delay}(a) 
($\tau<0$) can be fitted by the two-exponential form. However, 
it is difficult to fit the $+$ve delay data 
[Fig.~\ref{delay}(b)] this way.

To compare experiments with theory, coupled Maxwell-Bloch equations 
have to be solved~\cite{stroucken,jahnke} with appropriate boundary 
conditions and many body Coulomb effects to obtain the PPDR signal 
for layered semiconductor heterostructures as used in our experiments. 
However, the following simplified phenomenological approach may be 
useful in understanding the origin of the PPDR signal as a starting 
step. Here, the MQW structure is replaced by an effective single 
layer of exciton resonances. Neglecting the effects of multiple 
reflections within the barrier and well regions and within the cap 
layer and taking the same background refractive index for the 
barriers and the QWs as a first approximation, the reflectivity $R$ 
for the model sample structure (Fig.~\ref{expt}) can be written as

\begin{equation} \label{rfield}
R \propto \left|r_{0}\mathcal{E}_{2}+\mathcal{E}_{QW}e^{i\phi}\right|^{2},
\end{equation}

where only terms linear in the reflection coefficient $r_{0}$ ($<1$) of 
the cap layer on top of the QWs are retained. Here $\mathcal{E}_{2}$ 
is the amplitude of the probe electric field incident on the MQW 
and $\mathcal{E}_{QW}$ is that of the field emitted from the MQW 
layer. The phase factor $\phi=2dn\Omega/c$ arises from propagation 
delay for the probe pulse in the region between the vacuum-sample 
and top layer-MQW interfaces, $n$ is the refractive index of the cap 
layer, $c$ is the speed of light in vacuum and $\Omega$ is the 
circular frequency of the incident light. The PPDR signal $\Delta R$ 
is defined as $R_{\text{on}}-R_{\text{off}}$ where $R_{\text{on}}$ 
($R_{\text{off}}$) is obtained from Eq.~\eqref{rfield} when the QWs 
are excited (not excited) by the pump pulse.  In calculating
$R_{\text{on}}$ we assume that the pump pulse causes modulation
$\Delta \mathcal{E}_{QW}$ in $\mathcal{E}_{QW}$ but $r_{0}$
remains  unaffected.

For a thin sample with negligible depletion of incident pulse
intensities, and using slowly varying envelope approximation 
in time, it can be shown~\cite{bloembergen,stroucken} that
$\mathcal{E}_{QW}$ in absence of pump pulses is proportional 
to $i\mathcal{P}_{2}^{(1)}$ and the modulation 
$\Delta\mathcal{E}_{QW}$ caused by the pump pulse is proportional 
to $i\mathcal{P}^{N\!L}$, where $\mathcal{P}_{2}^{(1)}$ and 
$\mathcal{P}^{N\!L}$ are the amplitudes of the linear polarization 
induced by the probe pulse and nonlinear polarization induced by 
the pump and probe pulses, respectively. The lowest order term 
in $\mathcal{P}^{N\!L}$ is the third order polarization 
$\mathcal{P}^{(3)}$ which has products of the three amplitudes 
$\mathcal{E}_{2}$, $\mathcal{E}_{1}$ and $\mathcal{E}_{1}^{*}$ 
in the present case ($\mathcal{E}_{1}$ is the amplitude of the 
pump electric field incident on the MQW).

Keeping only the lowest order term in $\mathcal{P}^{N\!L}$, we get 
the TR-PPDR signal as

\begin{equation} \label{drtr}
\begin{split}
\Delta R(t, \tau) \propto &-\im \; [\mathcal{E}_{2}^{*} (t) \,
\mathcal{P}^{(3)} (t, \tau) e^{i \phi}] \\ & + C \re \;
[\mathcal{P}_{2}^{* \,(1)} (t) \, \mathcal{P}^{(3)}(t, \tau)],
\end{split}
\end{equation}

where $t$ is the real time and $C=2\pi\Omega n/cr_{0}$. Similarly, 
the SR-PPDR signal is given by

\begin{equation} \label{drsr}
\begin{split}
\Delta R(\omega, \tau) \propto &-\im\;[\mathcal{E}_{2}^{*}(\omega)
\, \mathcal{P}^{(3)}(\omega, \tau) \, e^{i \phi}] \\ & + C \re \;
[\mathcal{P}_{2}^{* \,(1)} (\omega)\,\mathcal{P}^{(3)}(\omega, \tau)],
\end{split}
\end{equation}

where  $\mathcal{E}_{2}(\omega)$, $\mathcal{P}_{2}^{(1)}(\omega)$,
and $\mathcal{P}^{(3)}(\omega)$ are the Fourier transforms of
$\mathcal{E}_{2}(t)$, $\mathcal{P}_{2}^{(1)}(t)$, and
$\mathcal{P}^{(3)}(t)$ respectively. Eqs.~\eqref{drtr} and 
\eqref{drsr} apply also for the PPDT case when the phase term 
$\phi$ is set to zero and $r_{0}$ in $C$ is replaced by unity.
The literature reports which are mostly concerned with SR-PPD 
measurements usually ignore~\cite{sokoloff,lindberg,koch} the 
second term in Eq.~\eqref{drsr}. In our investigation of the 
SR-PPDR case~\cite{bipul}, it was found that the second term is 
indeed smaller than the first but is not negligible. We wish to 
emphasize here that unlike the SR-PPD case, the second term is 
very crucial for the TR-PPDR (and TR-PPDT) measurements. In fact, 
within the above model, it is easy to show that the TR-PPDR 
measurements can not be explained without the second term in 
Eq.~\eqref{drtr}.

\begin{figure}[h]
\includegraphics[clip,width=6.5cm]{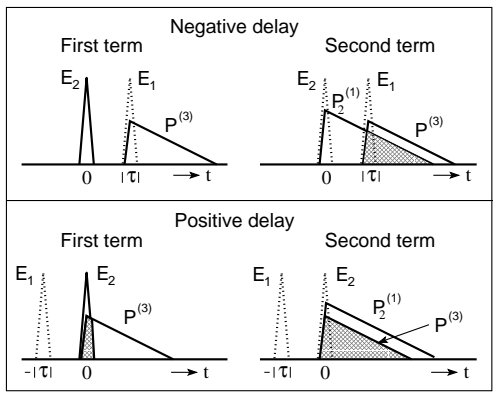}
\caption{ \label{schem}
Schematic representation of the origin of the TR-PPDR signal due 
to the two terms in Eq.~\eqref{drtr} at $-$ve (upper panel) and 
$+$ve (lower panel) delay. The shaded regions indicate the time 
domain in which the TR-PPDR signal arises due to overlap of 
$\mathcal{P}^{(3)}(t)$ in $t$ with $\mathcal{E}_{2}(t)$ (first 
term)  and $\mathcal{P}^{(1)}_{2}(t)$ (second term).
}
\end{figure}

Before evaluating $\Delta R(t,\tau)$, we may examine Eq.~\eqref{drtr} 
to understand qualitatively how the TR-PPDR signal arises. For 
this, we refer to the schematic pictures of Fig.~\ref{schem}. Let 
us assume that the probe and pump pulses are $\delta$-like in time.
Consider the $-$ve delay case. The probe pulse is taken to arrive 
on the sample at $t=0$ and the pump pulse later at $t=|\tau|$.  
$\mathcal{P}^{(3)}(t,\tau)$ is nonzero (and decaying with $t$) only 
for $t \ge |\tau|$.  In the first term of Eq.~\eqref{drtr}
$\mathcal{E}_{2}(t)$ is a $\delta$-like function at $t=0$. There 
is no overlap between $\mathcal{E}_{2}(t)$ and  
$\mathcal{P}^{(3)}(t,\tau)$ and hence no TR-PPDR signal is 
expected [Fig.~\ref{schem}~(upper panel)] from the first term.
However in the second term, $\mathcal{P}_{2}^{(1)}(t)$ is nonzero 
(and decaying with $t$) for $t\ge 0$. If $|\tau|$ is less than the 
exciton dephasing time, then there is an overlap in $t$ between 
$\mathcal{P}_{2}^{(1)}(t)$ and $\mathcal{P}^{(3)}(t,\tau)$ for 
$t \ge |\tau|$ and we get nonzero TR-PPDR signal 
[Fig.~\ref{schem}~(upper panel)] due to the second term in
Eq.~\eqref{drtr}. In the case of $+$ve delay, the pump pulse 
arrives on the sample at $t=-|\tau|$ and the probe pulse 
at $t=0$.  $\mathcal{P}^{(3)}(t,\tau)$ is nonzero only for 
$t \ge 0$ and decays with $t$. There is an overlap between
$\mathcal{E}_{2}(t)$ and  $\mathcal{P}^{(3)}(t,\tau)$ at 
$t=0$ and between $\mathcal{P}_{2}^{(1)}(t)$ and 
$\mathcal{P}^{(3)}(t,\tau)$ for $t \ge 0$. Hence for $+$ve delay, 
the first term contributes only at $t=0$ and the second term 
contributes for  $t \ge 0$ [Fig~\ref{schem}~(lower panel)]. In 
view of the above, observation of nonzero TR-PPDR signal (and in 
fact TR-PPDT signal as well) at $t>0$ for both $+$ve and $-$ve
$\tau$ in Figs.~\ref{delay} and \ref{int} can not be explained if 
only the first term in  Eq.~\eqref{drtr} is considered.

Furthermore, it is obvious from the $t$ ordering of 
$\mathcal{E}_{2}(t)$ and  $\mathcal{P}^{(3)}(t,\tau)$ that the 
time integral of the first term in Eq.~\eqref{drtr} vanishes
for $\tau<0$, giving no TI-PPDR signal. Thus, once again, the 
observed nonzero TI-PPDR signal for $-$ve $\tau$ in 
Fig.~\ref{expt}(c) can arise only from the $t$ integration of 
the second term in Eq.~\eqref{drtr} (which does not vanish). 
Equivalence of TI- and SI-PPDR signal can be established by 
using Parseval's theorem for Eqs.~\eqref{drtr} and \eqref{drsr}. 
This immediately shows that the first term in Eq.~\eqref{drsr} 
will lead to zero SI-PPDR signal, though it contributes to 
SR-PPDR measurements.

\begin{figure}[h]
\includegraphics[clip,width=7.0cm]{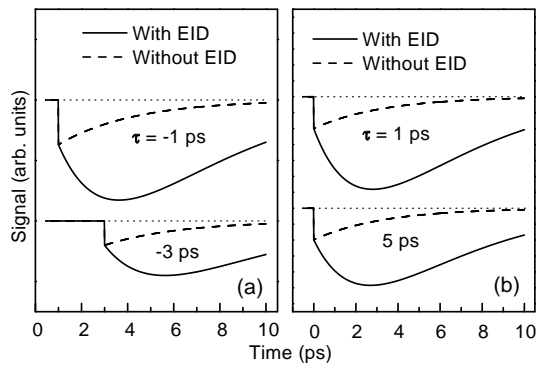}
\caption{ \label{sim}
Simulation of TR-PPDR signal using Eq.~\eqref{drtr} calculated 
within optical Bloch equation for $\delta$-like electric fields
for $-$ve (a) and $+$ve (b) delays, highlighting the role of 
excitation induced dephasing (EID). The signal is $-$ve and 
the curves for different $\tau$ are shifted along the vertical 
axis (linear scale), with the zero of the signal shown as dotted 
lines.
}
\end{figure}

We now assess how far a  calculation of $\mathcal{P}_{2}^{(1)}(t)$ 
and $\mathcal{P}^{(3)}(t, \tau)$ in Eq.~\eqref{drtr} based on a 
perturbative solution of the coupled optical Bloch equations for a 
non-interacting two-level assembly  up to third order in electric 
fields can reproduce the features shown by the experimental data. 
Many body effects like local fields (LF), excitation induced 
dephasing (EID) and resonance energy shift (ES) are included 
phenomenologically~\cite{bipul,wegener,wang} 
in this simplified approach. Although not done here, the 
coexcitation of hh and lh excitons can be easily included by 
modeling the excitonic assembly as a three-level 
system~\cite{srink,bipul}.  $\delta$-like pulses are assumed for 
computational ease although finite pulse widths can also be used.
We calculate the time evolution of the PPDR signal 
[$\Delta R(t,\tau)$] for both $-$ve and $+$ve $\tau$ 
(Fig.~\ref{sim}). Since the first term in Eq.~\eqref{drtr} 
contributes only for $+$ve delay at $t=0$ in a time domain of 
the size of the pulse width,   we display in Fig.~\ref{sim} only 
the calculation of the second term in Eq.~\eqref{drtr}. We find 
that only EID contributes to $\Delta R(t,\tau)$ apart from the 
phase space filling related nonlinearity. The peak behavior seen 
in Fig.~\ref{sim} is in fact a consequence of EID in this model. 
The LF and ES related terms lead the EID term by a phase of 
$\pi/2$ and do not contribute to the TR-PPDR. (However, they do 
contribute to the SR-PPDR signal together with EID~\cite{bipul}.)
In this model, the signal peak position in $t$ is essentially
determined by an exciton density independent dephasing rate.
The EID introduced in the Bloch equations does not lead to density
dependent decay of $\mathcal{P}^{(3)}(t, \tau)$. This is  
consistent with the data of Fig.~\ref{int}. Also, the model 
does lead to a delayed rise of the signal as observed in
Fig.~\ref{delay}. Within this theory, the $t$ integration of 
the second term in Eq.~\eqref{drtr}
produces a nonzero signal for $\tau<0$, with the behavior
$\sim \exp(2\tau/T_{2})$, $T_{2}$ being the exciton
dephasing time. This agrees with the data plotted in
Fig.~\ref{expt}(c), giving $T_{2} \approx 4$~ps, which is 
close to the value obtained from FWM measurements on this 
sample. 

However, some features in the data are not explained by the 
simple theory. For example, the origin of the different rising 
behavior in Figs.~\ref{delay} and \ref{int} for $-$ve and $+$ve 
$\tau$ is not clear. Furthermore, although the TR-PPDR signal decay 
appears to be not very sensitive to increase in the pump intensity 
(Fig.~\ref{int}), the rise is. This is not accounted for in the 
above theory. Introduction of a phenomenological dependence of 
the dephasing time for the probe induced exciton polarization on 
pump generated excitation density may be considered, in analogy 
with the approach followed in the literature for FWM~\cite{wang}. 
This will indeed cause a faster rise of the TR-PPDR signal in $t$ 
with increased pump intensity ($I_{1}$). However, this will also 
lead to a faster decay in $t$ of the signal with increasing 
$I_{1}$, which is not evident in Fig.~\ref{int}. A quantitative 
understanding of the $t$ and $I_{1}$ dependence exhibited by the 
data of Figs.~\ref{delay} and \ref{int} may be possible when an 
appropriate theory of the nonlinear optical interactions of the 
excitonic polarization, populations and propagating electric 
fields, that also takes into account the MQW nature of the sample
and higher order Coulomb interactions, is applied. This is beyond 
the scope of this paper.

Among the important effects that may play some role in the TR-PPDR 
measurements is the following. It has been 
suggested~\cite{stroucken,baumberg,prineas} that 
radiative coupling of excitations in individual QWs 
in a MQW structure may give rise to a few bright polaritonic modes 
which reradiate in the reflection direction more efficiently than 
other modes. One consequence of this theory is that the tail of 
the emission from the linear polarization may show a finite rise 
time and beats due to polaritonic interference. Indeed, it was 
deduced recently~\cite{baumberg} from measurements of time 
integrated reflectivity of a pair of phase locked pulses using an 
interferometric technique that the excitonic reflection in GaAs 
MQWs shows a finite rise time at low excitation intensities.

\begin{figure}[h]
\includegraphics[clip,width=4.0cm]{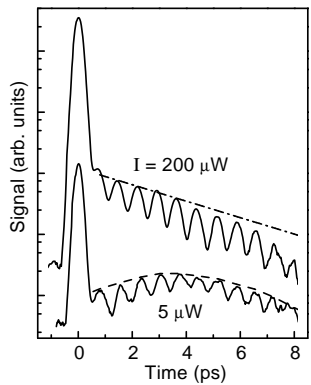}
\caption{ \label{ppr}
Semilogarithmic plot showing the time profile of the coherent 
emission along the probe reflection direction (no pump pulse) 
for two average intensities ($I=200$ and 5~$\mu$W) of the 
incident pulse. Data at two $I$ are shifted vertically. The 
dash-dotted line is an exponential fit, the dashed curve is 
drawn to guide the eye.
}
\end{figure}

To investigate this aspect in our case, the temporal profile of 
the reflected probe pulse is measured in absence of a pump pulse 
(Fig.~\ref{ppr}). The strong peak at $t=0$ is essentially due to 
reflection of the probe pulse at the vacuum-sample interface. 
The oscillations seen are quantum beats due to coherent excitation 
of hh and lh excitons. The exciton polarization emits coherently 
for several ps. The emission is shown for two values of the average 
intensity $I$ of the incident pulse, namely $I=200$ and  5~$\mu$W.
A simple estimate gives an exciton density of 
$8\times 10^{6}$~cm$^{-2}$ per $\mu$W per QW. At low intensities, 
the emission shows a finite rise time, the signal peaking at about 
3~ps. This behavior is consistent with the possibility that 
polaritonic interference has an important effect on the dynamics 
of excitonic polarization in MQWs and may influence TR-PPDR 
measurements. This feature is not seen for the high 
intensity case, possibly due to reduced photon induced coupling
between the excitons in the individual QWs caused by exciton dephasing. 
However, even in this case, the decay of the signal shows a departure 
from a simple exponential behavior, as indicated in Fig.~\ref{ppr}.
We should mention that the $t$ dependence seen in Fig.~\ref{ppr} is 
different from the nonexponential decay that may be caused by exciton 
inhomogeneous broadening when it is comparable to or more than the 
homogeneous broadening.

To conclude, the TR-PPDR results presented here reveal new features
related to the time dynamics of excitonic polarization in MQWs and 
the dependence of pump modulation effects on pump intensity, 
pump-probe relative polarization and delay. A nonlinear optical 
contribution arising from interference of excitonic linear and 
nonlinear polarization, which was not considered earlier in the 
literature, was shown to be essential for understanding the 
TR-PPDR measurements.

We are grateful to L.~N.~Pfeiffer and J.~Shah for the MQW sample.

\end{document}